\documentclass[10pt,conference]{IEEEtran}
\IEEEoverridecommandlockouts

\usepackage{cite}
\usepackage{amsmath,amssymb,amsfonts}
\usepackage{graphicx}
\usepackage{textcomp}
\usepackage{xcolor}
\usepackage{xspace}
\usepackage{algorithm}
\usepackage{algpseudocode}
\usepackage{amsmath}
\usepackage{dsfont}
\usepackage{xcolor}
\usepackage{tikz}
\usepackage{color,colortbl}
\usepackage{tcolorbox}
\usepackage{multirow}
\usepackage{enumerate}
\usepackage{makecell}
\usepackage{arydshln}
\usepackage{pifont}
\usepackage{mdframed}
\usepackage{booktabs} 
\usepackage{url}
\usepackage{stfloats}

\definecolor{deepred}{rgb}{0.55, 0, 0}
\definecolor{blueish}{RGB}{250, 250, 255}
\definecolor{greenish}{RGB}{200, 255, 200}
\definecolor{redish}{RGB}{255, 200, 200}
\definecolor{highlight}{RGB}{175, 255, 100}
\definecolor{darkred}{RGB}{139, 0, 0}
\definecolor{gray95}{gray}{0.05}
\definecolor{rowgray}{RGB}{224, 224, 224}

\newmdenv[
    tikzsetting= {fill=blueish},
    skipabove=0.33em,
    skipbelow=0.33em,
    linewidth=1pt,
    innerleftmargin=4pt,
    innerrightmargin=4pt,
    innertopmargin=2pt,
    innerbottommargin=2pt,
    linecolor=gray95,
    roundcorner=2pt, 
    shadowsize=4pt,
    shadowcolor=gray95
]{answerbox}

\newenvironment{result}
{\begin{answerbox}}
{\end{answerbox}}

\newcommand{\ourapproach}{{\sc RoCode}\xspace}
\newcommand{\ourapproachbf}{{\sc \textbf{RoCode}}\xspace}

\def\BibTeX{{\rm B\kern-.05em{\sc i\kern-.025em b}\kern-.08em
    T\kern-.1667em\lower.7ex\hbox{E}\kern-.125emX}}

\begin{document}

\title{ROCODE: Integrating Backtracking Mechanism and Program Analysis in Large Language Models for Code Generation}

\author{
\IEEEauthorblockN{Xue Jiang, Yihong Dong}
\IEEEauthorblockA{\textit{Key Lab of High Confidence Software} \\
\textit{Technology, MoE (Peking University)} \\
Beijing, China \\
\{jiangxue, dongyh\}@stu.pku.edu.cn}
\and
\IEEEauthorblockN{Yongding Tao}
\IEEEauthorblockA{\textit{University of Electronic Science and } \\
\textit{Technology of China}\\
Chengdu, China \\
yongd.tao@gmail.com}
\and
\IEEEauthorblockN{Huanyu Liu}
\IEEEauthorblockA{\textit{Key Lab of High Confidence Software} \\
\textit{Technology, MoE (Peking University)} \\
Beijing, China \\
huanyuliu@stu.pku.edu.cn}
\and
\IEEEauthorblockN{Zhi Jin}
\IEEEauthorblockA{\textit{Key Lab of High Confidence Software} \\
\textit{Technology, MoE (Peking University)} \\
Beijing, China \\
zhijin@pku.edu.cn}
\and
\IEEEauthorblockN{Wenpin Jiao}
\IEEEauthorblockA{\textit{Key Lab of High Confidence Software} \\
\textit{Technology, MoE (Peking University)} \\
Beijing, China \\
jwp@sei.pku.edu.cn}
\and
\IEEEauthorblockN{Ge Li}
\IEEEauthorblockA{\textit{Key Lab of High Confidence Software} \\
\textit{Technology, MoE (Peking University)} \\
Beijing, China \\
lige@pku.edu.cn}
}

\maketitle

\begin{abstract}
Large language models (LLMs) have achieved impressive performance in code generation recently, offering programmers revolutionary assistance in software development.
However, due to the auto-regressive nature of LLMs, they are susceptible to error accumulation during code generation. Once an error is produced, LLMs can merely continue to generate the subsequent code conditioned on it, given their inability to adjust previous outputs. Existing LLM-based approaches typically consider post-revising after code generation, leading to the challenging resolution of accumulated errors and the significant wastage of resources. Ideally, LLMs should rollback and resolve the occurred error in time during code generation, rather than proceed on the basis of the error and wait for post-revising after generation. In this paper, we propose \ourapproachbf, which integrates the backtracking mechanism and program analysis into LLMs for code generation. Specifically, we employ program analysis to perform incremental error detection during the generation process. When an error is detected, the backtracking mechanism is triggered to priming rollback strategies and constraint regeneration, thereby eliminating the error early and ensuring continued generation on the correct basis. Experiments on multiple code generation benchmarks show that \ourapproachbf can significantly reduce the errors generated by LLMs, with a compilation pass rate of 99.1\%. The test pass rate is improved by up to 23.8\% compared to the best baseline approach. Compared to the post-revising baseline, the token cost is reduced by 19.3\%. Moreover, our approach is model-agnostic and achieves consistent improvements across nine representative LLMs.

\end{abstract}


\section{Introduction}
As modern software architectures continue to increase in size and complexity, the burden on developers to construct and maintain these systems has become substantial. Given that programs serve as the fundamental carriers of software functionality, the automation of their generation is of paramount importance. Code generation technology, which seeks to automatically produce programs that align with human intentions, has emerged as a focal area of interest within both academia and industry fields \cite{codex, Subtoken-TranX, CODEP, CodeLlama}. In recent years, large language models (LLMs) have rapidly advanced and achieved significant success in the domain of automated code generation \cite{codegen, incoder, self-planning, SEED, CR}. A well-known tool for code generation based on LLMs is Copilot \cite{copilot}, which has demonstrated its utility by generating code that can be accepted by more than 30\% of its users \cite{copilot_acceptance_rate}. 

\begin{figure}[th!]
    \centering
    \includegraphics[width=0.48\textwidth]{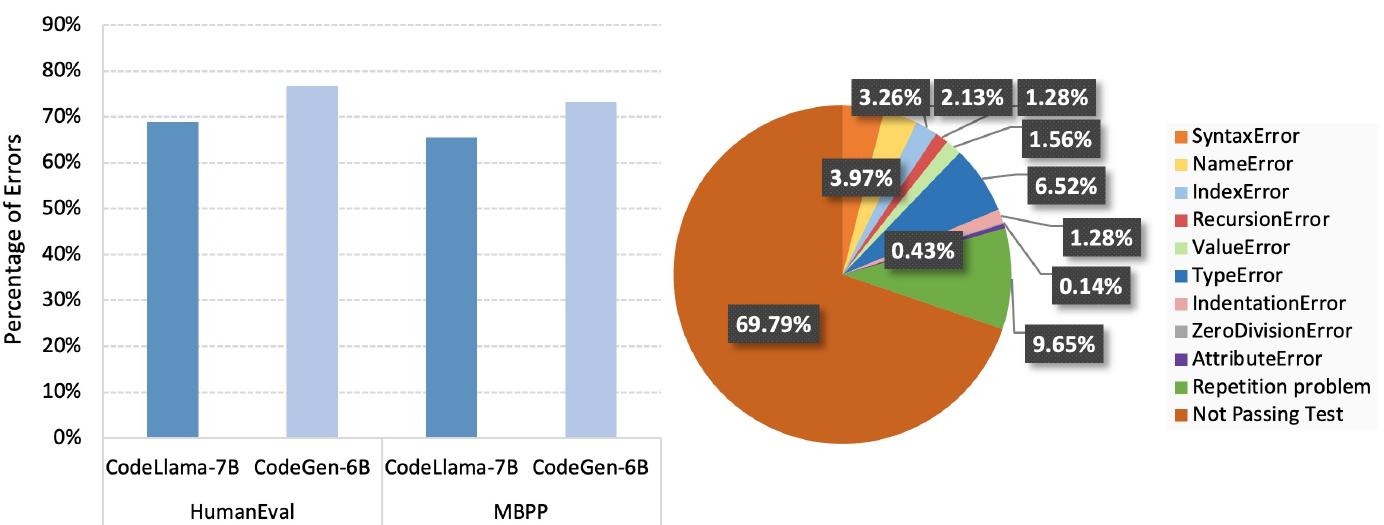}
    \caption{Statistics on the types of errors in code generated by LLM. The statistics are conducted based on the results generated by CodeLlama-7B and CodeGen-6B on HumanEval and MBPP benchmarks using greedy decoding.}
    \label{error_statistics}
\end{figure}

Typically, LLMs adopt an auto-regressive approach, where the output at each step is conditioned on the outputs of previous steps. Once an error occurs during the generation process at any step (for example, the selection of an inappropriate token due to hallucinations\footnote{The hallucination in code generation manifests as generated code that violates programming principles, resulting in code that cannot be compiled or executed, or that is inconsistent with user requirements or context, leading to failed tests \cite{liufang_Hallucinations}. Recent research has demonstrated that all computable LLMs cannot prevent themselves from hallucinating \cite{inevitable_hallucination}. } \cite{Hallucination_def_1}), this error will be included in the context of the subsequent steps. This phenomenon can cause errors to accumulate and amplify their impact, potentially causing the generated content to completely deviate from the expected path \cite{error_propagation_2, error_propagation_1}. Moreover, the generation process of LLMs differs significantly from the common practice of reviewing and adjusting existing code in human coding. In practice, developers are able to adjust their code whenever necessary based on its quality and its alignment with requirements, while LLMs can merely proceed based on the output generated so far and are unable to adjust previous outputs spontaneously. Recent studies \cite{Self-collaboration,self_refine,debug} have attempted to utilize the LLMs to revise their output after generation in a post-revising manner. However, this type of approach faces difficulties in revising the accumulated errors \cite{olausson2023self} and can result in resource wastage \cite{Self-collaboration}.

Ideally, through incorporating a backtracking mechanism into the generation process, we can expose potential errors early and resolve them, effectively preventing error propagation.
However, to effectively implement backtracking, three key issues ought to be addressed:
1) \textbf{When to roll back.} During the generation process, the rollback is triggered depending on when errors are detected. Error detection during the generation of LLMs should satisfy the following conditions. First, it must be capable of performing real-time checks on incomplete code; second, it is required to cover common errors produced by LLMs which are shown in Figure \ref{error_statistics}; finally, its running speed would be better to fast enough so as not to affect the efficiency of LLM significantly.
2) \textbf{Where to roll back to.} Simply rolling back to the last error-free state of the generated code usually does not address the issue. We should identify the initial decision point that caused the error and roll back to that point. Determining the rollback point is a complex decision-making process because the meaning and behavior of the erroneous code depend not only on itself, but also on interactions with preceding code, which are influenced by factors such as variable scopes, state dependencies, and logical dependencies within the program. 
3) \textbf{How to avoid previous errors.} 
After the rollback, the key task during regeneration is to prevent the recurrence of previous errors. However, completely prohibiting the LLMs from generating previously erroneous code may inadvertently block benign tokens. Thus, it is essential to impose appropriate constraints on the regeneration process.

To address the preceding three issues, we first implement incremental error detection using program analysis, which enables the examination of incomplete code to identify potential errors. Compilers can be used not only in code transformation for execution but also as an effective tool for program analysis. It is capable of performing numerous key and common analyses such as syntax parsing, type checking, and dependency analysis, and they have been optimized for speed over many years. Moreover, by using compilers, we can design new analyses for specific errors in LLMs' generated code, such as checking for code repetition problems. Second, for determining rollback points, program analysis serves as an external inspection during the generation of LLMs, providing essential error information. However, this error information may not directly pinpoint the root cause of the error. In contrast, the inherent uncertainty of LLMs is proven to be usable for self-assessment during generation \cite{semantic_entropy_nature, Semantic_Uncertainty, Semantic_Entropy_Probes}, which can aid in tracing the root cause of errors. Therefore, combining these two sources of information facilitates determining rollback points. Third, in regeneration with constraints, we decay the generation probability of the paths leading to error progressively. Moreover, by modeling the entire generation process with tree structures, it is feasible to comprehensively account for all historical errors and to effectively superimpose penalties for them.

In this paper, we propose \ourapproach, a novel code generation approach that integrates backtracking mechanism and program analysis to LLMs. The core of our approach -- the backtracking mechanism detects errors in real-time, rolls back, and regenerates with constraints during the generation process of LLMs, thus preventing error accumulation and enhancing the performance and efficiency of code generation. Specifically, we employ program analysis to perform incremental error detection during the code generation process to discover errors timely. Based on the results of program analysis and the observation of uncertainty in the generation of LLMs, we design a series of rollback strategies to determine the rollback point. To constrain the process of regeneration, we strategically penalize the likelihood of tokens that have contributed to previous errors. Further, given that the introduction of rollback and regeneration makes code generation no longer follow a linear path, we use a Trie Tree to model the whole generation process of \ourapproach. Importantly, our approach is model-agnostic, and requires no additional training.

Our experimental results demonstrate that \ourapproach consistently outperforms all baselines across six code generation benchmarks. \ourapproach achieves a compilation success rate of 99.1\% and surpasses the best-performing baseline by 23.8\% in pass rate. To further demonstrate its utility, we apply \ourapproach to multilingual code generation tasks and achieve a relative improvement of 34.2\% in pass rate. We also explore generalizability of \ourapproach across various LLMs, revealing significant enhancements in the performance of both general LLMs and code LLMs, with an average improvement of 18.2\% in pass rate. In terms of cost and performance, \ourapproach reduces token costs by 19.3\%, compared to the Post-revising approach. Furthermore, the ablation studies reveal that incremental error detection, rollback strategies, and constraint regeneration in \ourapproach all contribute to performance improvement. To the best of our knowledge, this work is the first to introduce and implement the rollback approach for code generation during the decoding process in LLMs\footnotemark. \footnotetext{Code is available at \url{https://github.com/jiangxxxue/ROCODE}.}

\section{Methodology}

\subsection{Overview}
\label{Overview}

For a code generation task, given the requirement $x$, we propose to perform \ourapproach for LLMs to generate code $y$. \ourapproach consists of three key steps:

\begin{itemize}
    \item \textbf{Incremental Error Detection} focuses on continuously checking the generated code during the generation process to discover errors early. By implementing program analysis, we can detect potential errors in the generated code such as compile errors and runtime errors.

    \item \textbf{Strategic Rollback}, upon detecting an error, rolls back the generated code to an earlier error-free state. In this step, we design a series of specific rollback strategies to determine the rollback point.

    \item \textbf{Constraint Regeneration} formulates error-related constraints and combines them with the LLM decoding process to prevent previous errors from happening again. 
    This step involves strategically penalizing the likelihood of the generated tokens that contributed to the errors.
\end{itemize}

To track the code generation progression of \ourapproach, we employ the structure of Trie Tree and and integrate operations of incremental error detection, strategic rollback, and constraint generation within the Trie Tree. 
This structure helps organize non-linear, tree-like code generation trajectories, allowing us to efficiently handle multiple rollback and regeneration cycles.

\subsection{Incremental Error Detection}
\label{Error Detection}

Considering LLMs are generated in an auto-regressive way, once an error occurs during the generation process, the LLMs will continue to generate content on the basis of the errors, leading to the propagation of errors. Since the occurred errors are inevitable in the final outputs, the subsequent generation derived from these erroneous contents can be almost considered redundant. Therefore, we employ incremental error detection to detect errors during generation in a timely manner and substantially reduce the cost of long rollbacks. 

Incremental error detection employs the program analysis tool to incrementally detect errors following the generation of each detectable unit. Specifically, we use the statement as the smallest unit for detection, each representing the smallest code unit with independent functionality. The LLM $\mathcal{M}$ incrementally generates these statements step by step. Upon completion of each statement, we employ the program analysis tool $C$ to conduct error detection. This process can be formulated as follows: 
\begin{align}
    s_i &= \mathcal{M}(x, S_{:i-1}), \\
    e_i &= \mathcal{C}(S_{:i-1} \parallel s_i),
    \label{ei}
\end{align}
where $s_i$ is the $i$-th statement generated by $\mathcal{M}$, $S_{:i-1} = [s_0, s_1, \ldots, s_{i-1}]$, $\parallel$ denotes the concatenation of statements, and $e_i$ is the report of incremental error detection for $s_i$, which is defined as:
\begin{align}
    e_i = \{\text{result}, \text{type}, \text{lineno}, \text{offset} \},
\end{align}
where `result' indicates the detection result of whether the generated code passes, with possible values being \{success, failure\}. If `result' is `success', the remaining items are not applicable, otherwise $e_i$ returns `failure' along with its `type', `lineno', and `offset'. Among them, `type' indicates the type of error detected, `lineno' represents the line number where the error occurs, and `offset' represents the specific position of error within the `lineno'.

The program analysis tool determines the types of errors that can be detected during error detection. There are various tools designed for different errors and programming languages. In this paper, we choose the compiler to support our program analysis. The reason is that the compiler integrates some key and mature analysis techniques, which can effectively detect errors commonly found in LLMs (shown in Fig. \ref{error_statistics}). Moreover, compilers support almost all programming languages and run fast. In the generation process, we use a compiler either without executing or with executing test input to check the generated code. Without executing, we can check for syntax errors, type mismatches, declaration errors, scope errors, and linking errors. With executing test input, we can further increase checks for runtime errors, including timeouts, recursion errors, division by zero errors, memory access errors, index out-of-bounds errors, and resource not found errors. Moreover, in practical scenarios, developers usually have access to publicly available test cases to better understand and validate requirements, we take this part into account. Once the code is completely generated, we execute the complete set of test cases (if available), which include both input and output, to thoroughly verify the program's logic.

Furthermore, we observe that repeat patterns problem occurs during the code generation process, characterized by the repetitive output of the same syntactic structure but meaningless code constructs like `if-elif-elif...', `print', etc., resulting in failure to generate a termination symbol ($EOS$) \cite{liufang_Hallucinations}. This problem typically does not result in syntax and compile errors during generation, but it can significantly affect the semantics of the generated code, thereby introducing potential logic errors. Therefore, we design an additional analysis to detect repeat patterns problem. Specifically, we utilize the syntax parsing module of the compiler to extract the syntactic structure and identify repetitive patterns. According to Abstract Syntax Definition Language (ASDL) \cite{ASDL}, if the same type of \textit{stmt} appears consecutively more than a specified number of times, it is considered as an error with the error report including `type' as repetition and `lineno' as the line number where the first repeated \textit{stmt} occurs.

\begin{figure*}[th!]
    \centering
    \includegraphics[width=0.9\textwidth]{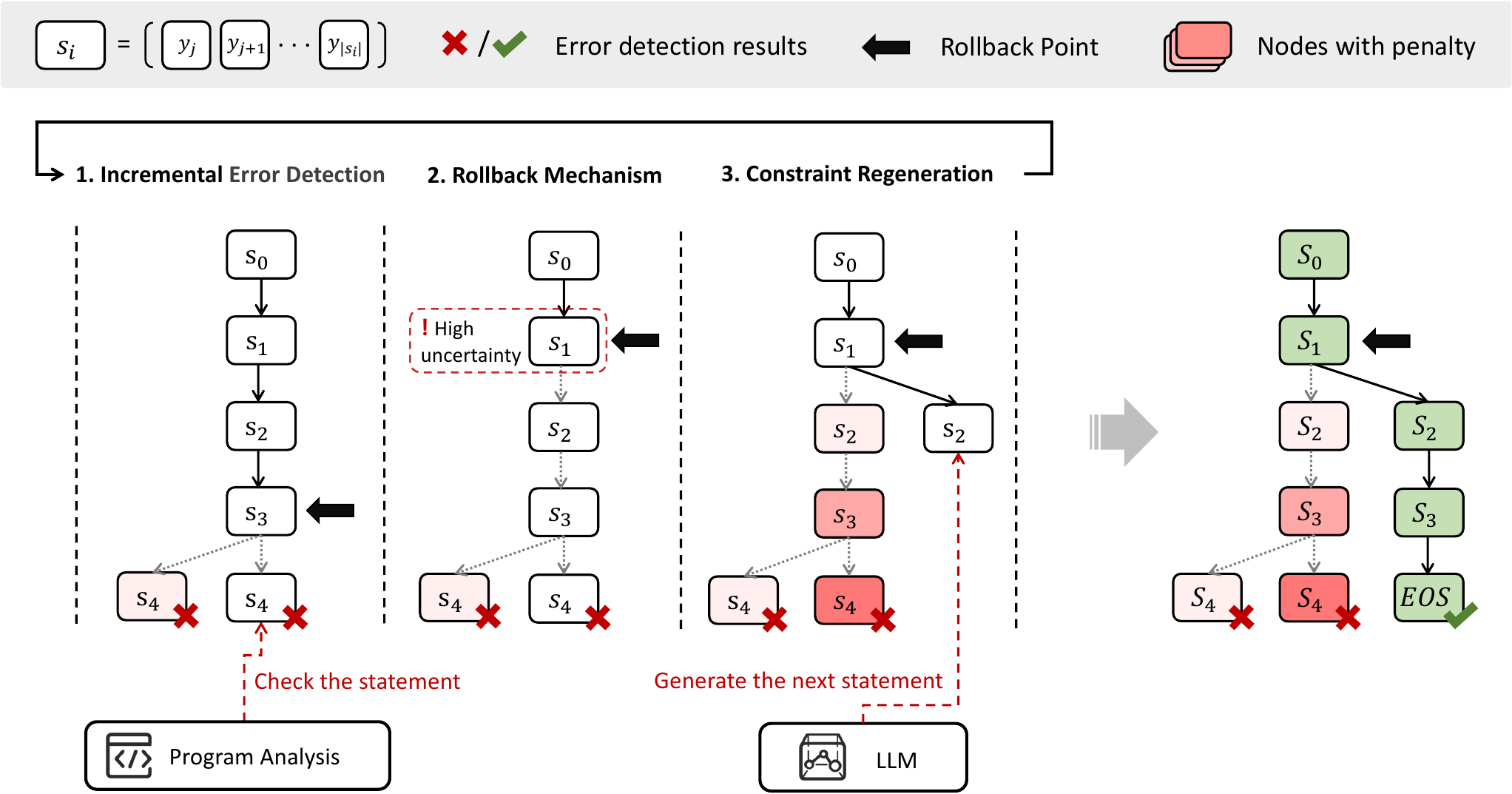}
    \caption{The Overview of \ourapproach with Trie Tree.}
    \label{trie_tree}
\end{figure*}

\subsection{Strategic Rollback}
\label{Strategic Rollback}

When the error is detected, it is necessary to undo a part of the previously generated code to rectify the issue. To identify the specific point that requires to roll back to, we design a series of strategies to determine the rollback point.

Generally, for detected errors, 
incremental error detection can provide an error report including the location where the error occurs, offering an initial clue to resolve the error. Therefore, we first attempt to resolve the error by rolling back directly to this specific location. The rollback point $r$ is defined as a two-dimensional value:
\begin{equation}
    r_e = \left[e.\text{lineno}, e.\text{offset}\right],
    \label{re}
\end{equation}
where $e.\text{lineno}$ refers to the line number of error report $e$ and $e.\text{offset}$ represents the offset within that line, used to precisely locate the error.
However, merely reverting to $r_e$ may not be sufficient for some complex errors, since the root cause of these errors might not actually originate from the reported location instead of an early location. These errors usually involve dependencies and semantics. 

Given that LLMs are probabilistic models, i.e., LLMs predict the next token by calculating the conditional probability of each possible subsequent token given the preceding context, LLMs may exhibit high levels of uncertainty at certain points during the generation process, leading to fluctuating outputs \cite{semantic_entropy_nature, Semantic_Uncertainty, Semantic_Entropy_Probes}. Although high uncertainty increases the likelihood of erroneous decisions, it also facilitates the redistribution of probabilities to alter the output. Therefore, we can infer rollback points by analyzing the model's display of uncertainty. We can calculate the entropy at each position using the following formula:
\begin{equation}
    H_t = -\sum_{j=1}^{|V|} p(y_t = v_j \mid y_{<t}, x) \log p(y_t = v_j \mid y_{<t}, x),
    \label{ht}
\end{equation}
where $p(y_t = v_j \mid y_{<t}, x)$ denotes the probability of generating the $t$-th token $ y_t $ as $ v_j $ given the context $ x $ and previously generated tokens $ y_{<t} $. The summation $ \sum_{j=1}^{|V|} $ iterates over all possible tokens $ v_j $ in the vocabulary $ V $ that can be generated at this position. 
We roll back to the beginning of the statement containing the token with the highest entropy.
\begin{equation}
    t^* = \text{argmax}_{t \in [0, |y|]} H_t 
    \label{tstar}
\end{equation}
\begin{equation}
r_h = \left[\text{ConvertToLineno}(t^*, y), 0\right],
\end{equation}
where $\text{ConvertToLineno}$ is a function that converts the token position to the corresponding line number within the code, and $|y|$ denotes the length of generated code $y$. This strategy allows for an opportunity to refresh the most relevant context in the next generation, thereby avoiding the recurrence of this high-entropy point. The algorithm for the strategic rollback is shown in Algorithm \ref{algorithm1}.

\begin{algorithm}
\caption{Algorithm of Strategic Rollback}
\label{Rollback Algorithm}
\label{algorithm1}
\begin{algorithmic}[1]
\Require  Error Detection Reports $E = \{e_i\}_{1:n}$ and Generated Statements $S = \{s_i\}_{1:n}$.
\Ensure Rollback Point $r$.  
\State Assert $e_n.\text{result is `failure'}$.
\State Initialize $y \gets s_1 || \cdots || s_n$.
\If {$e_n.\text{lineno}$ and $e_n \neq e_{n-1}$} 
    \State $r \gets \left[e.\text{lineno}, e.\text{offset}\right]$.
\Else \Comment{$e_n.\text{lineno}$ is None or the error recurs.}
    \State $r \gets \left[\text{ConvertToLineno}(t^*, y), 0\right]$, where $t^*$ is computed via Eq. \eqref{ht} and Eq. \eqref{tstar}.
\EndIf
\State \Return $r$
\end{algorithmic}
\end{algorithm}

\subsection{Constraint Regeneration}
\label{Constraint Regeneration}

After error detection and rollback, we perform constraint regeneration to prevent the LLM from reproducing the same error. Constraint regeneration involves two parts: constructing constraints and integrating these constraints into the LLM's decoding process, thereby influencing the model's output behavior.

\textbf{Constraint Construction:}
We define constraints as penalties applied to the LLMs' output probabilities for previously generated erroneous code. In the process of code generation, the generated code can be considered as an output path of LLMs. To avoid the model generating incorrect paths, a naive approach is to set the probabilities of tokens on erroneous paths to zero, completely blocking those paths. However, this approach can penalize some benign tokens. Instead, we adopt a milder penalty approach, which applies an exponentially decaying penalty to each token from the point of error back to a rollback point,
\begin{equation}
    \text{PN}(v \mid y_{<t}) =
    \begin{cases}
    \lambda^{t-r}, & \text{if } v = y_t\text{,} \\
    1, & \text{otherwise,}
    \end{cases}
    \label{pn}
\end{equation}
where $\lambda$ is a decay factor between 0 and 1 and $t-r$ denotes the number of time steps from the rollback point $r$ to the current token $y_t$. 
This approach not only penalizes the tokens that directly cause errors but also applies lighter penalties to preceding tokens that may have indirectly contributed to the mistake, thereby preventing the model from repeating erroneous generation paths.

\textbf{Decoding with Constraints:} 
To avoid errors in the previous generation path, penalties are applied to the probability distribution of LLMs, adjusting the generation likelihood of each token. Then, this modified probability distribution is re-normalized. Specifically, for each token $y'_t$, its constrained probability distribution $ p_c(y'_t \mid y_{<t}) $ is given by:
\begin{equation}
    p_c(y'_t \mid y_{<t}) = \frac{p(y'_t \mid y_{<t}) \cdot \text{PN}(y'_t \mid y_{<t})}{\sum_{v} p(v \mid y_{<t}) \cdot \text{PN}(v \mid y_{<t})}
    \label{cpd}
\end{equation}

\subsection{Trie Tree Modeling}
\label{Trie Tree Modeling}
The generation process of \ourapproach involves rollbacks and regeneration, resulting in a non-linear structure. Therefore, we use Trie Tree to model the entire process, as illustrated in Figure \ref{trie_tree}. In the Trie Tree $ T = (U, E) $, each node $u \in U$ represents a generated token in this process, and each edge $ (u,v) $ indicates that the token sequence from the root node to node $u$ serve as the context to generate node $v$.

During the generation process of \ourapproach, as each statement is generated, its corresponding tokens are sequentially appended to the Trie Tree. This addition is immediately followed by incremental error detection. If an error is identified, the affected path within the tree is flagged and the strategic rollback is activated, identifying the precise node to revert to. Subsequently, all descendant nodes of this rollback point, representing the erroneous sequence, are used to impose constraints. This penalization process effectively discourages the regeneration of the same erroneous sequences during subsequent iterations of code generation. Each new, error-free statement is integrated into the tree as a distinct branch, aligning with existing paths that share a common prefix. This integration not only consolidates the tree structure but also accumulates the penalties associated with each erroneous path, reinforcing the deterrent against repeating past mistakes. It ensures that the generation process dynamically adapts, minimizing the recurrence of similar errors and optimizing code output over time.

Ultimately, each path from the root node to any terminal node represents an attempt at the generation process of \ourapproach, and the last path of Trie Tree represents the final generated code $y$. The pseudocode of \ourapproach during code generation is shown in Algorithm \ref{workflow}.

\begin{algorithm}
\caption{ The Pseudocode of \ourapproach} \label{workflow}
\begin{algorithmic}[1]

\Require Input Requirement $x$, LLM $\mathcal{M}$.
\Ensure Generated Code $y$.

\State Initialize Trie Tree $T \gets \emptyset$ and index $i \gets 0$.
\State Statement $s_{i} \gets \mathcal{M}(x)$
\State $T$.update\_stmt($s_i$).
\While{$s_{i}$ does not include EOS token}
    \State \textcolor{gray}{\textit{\# \textbf{Incremental Error Detection}}}
    \State $e_i \gets \mathcal{C}(T.\text{stmts})$ via Eq. \eqref{ei}.
    \State $T$.update\_report($e_i$). 
    \State \textcolor{gray}{\textit{\# \textbf{Strategic Rollback}}}
    \If{$e_i.\text{result is `failure'}$} 
        \State $r \gets \text{RollBack}(T.\text{stmts}, T.\text{reports})$ via Alg. \ref{Rollback Algorithm}. 
        \State $T$.rollback\_to($r$). 
    \EndIf
    \State \textcolor{gray}{\textit{\# \textbf{Constraint Generation}}}
    \State $T$.update\_pn($T$.\text{stmts}, $r$)  via Eq. \eqref{pn}.
    \State Sample $s_{i+1} \gets \mathcal{M}(x, T.\text{stmts}, T.\text{pn})$ via Eq. \eqref{cpd}.
    \State $i \gets i+1$.
    \State $T$.update\_stmt($s_i$).
\EndWhile

\State \Return $T.\text{get\_final\_gen\_code}()$
\end{algorithmic}
\end{algorithm}

\section{Evaluation}
\ourapproach aims to effectively prevent error accumulation in the code generation process of LLMs and improve the quality of generated code by integrating backtracking mechanism and program analysis into LLMs.
In this section, we present extensive experiments that span six representative code generation benchmarks, two program languages, and nine LLMs of varying series or sizes. We aim to investigate six research questions: 
\begin{itemize}
\item RQ1: How does \ourapproach perform compared to baseline approaches on code generation benchmarks?
\item RQ2: How effective is \ourapproach in improving LLMs in code generation tasks across different programming languages?
\item RQ3: How does \ourapproach perform when applied to different LLMs?
\item RQ4: How about the cost and efficiency of \ourapproach?
\item RQ5: How does each component of \ourapproach contribute to the effectiveness?
\item RQ6: How does the hyperparameter decay factor affect the effectiveness of \ourapproach?
\end{itemize}

\subsection{Evaluation Setup}

\noindent \textit{\textbf{1) Benchmark:}}
We perform a comprehensive evaluation on six code generation benchmarks to demonstrate the superiority and generality of \ourapproach. 

\textbf{HumanEval} \cite{codex} consists of 164 handwritten programming tasks, proposed by OpenAI. Each task includes a function signature, a requirement, use cases, a function body, and several unit tests (average of 8 per task). We use the use cases as public test cases for our approach and baseline approaches, while unit tests are used as private test cases for evaluation.

\textbf{MBPP} \cite{mbpp} contains 974 Python programming tasks, covering programming fundamentals, standard library functionality, and more. The MBPP dataset does not specify public vs. private test cases. Following previous work \cite{MBR}, we use one input of the test cases for all baseline approaches and do not involve any ground-truth test case output.

\textbf{CodeForces2305} \cite{dong-etal-2024-generalization} comprises 90 of the competition-level programming problems collected from the CodeForces website. On average, each problem is accompanied by three public test cases and three private test cases. These problems are created after May 2023, which is after the training data cutoff of most LLMs, such as CodeLlama \cite{CodeLlama} and CodeGen \cite{codegen}, mitigating the impact of data contamination on evaluation.
    
\textbf{HumanEval-ET} and \textbf{MBPP-ET} \cite{CodeScore} are expanded versions of HumanEval and MBPP with over 100 additional test cases per task. This updated version includes edge test cases that enhance the soundness of code evaluation compared to the original benchmark.  

\textbf{HumanEval-CPP} \cite{codegeex} is constructed based on the HumanEval benchmark to evaluate the code generation ability of LLMs on C++ programming language.
\\
\\
\noindent \textit{\textbf{2) Baselines:}}
Our approach works on the decoding phase of LLMs that does not require modification and training of the model. We use the three most common decoding approaches of LLMs and set them as baselines. Specifically,

\textbf{Temperature Sampling} \cite{caccia2019language} controls the randomness of the token selection process—higher temperatures $T$ lead to more uniform distributions, while lower temperatures $T$ make high-probability tokens even more likely.
    \begin{equation}
        P'(w) = \frac{\exp(\log(P(w \mid w_{<t})) / T)}{\sum_{w'} \exp(\log(P(w' \mid w_{<t})) / T)},
    \end{equation}
    when $T$ is 0, $P'(w)$ is equivalent to $\mathds{1}(w=\arg\max_w P(w \mid w_{<t}))$, which means greedy sampling.

\textbf {Topk Sampling} \cite{topk} limits the next-word selection to the top k most likely candidates as determined by the model.
\begin{equation}
   P'(w) = 
   \begin{cases} 
   P(w \mid w_{<t}) & \text{if } w \in \text{Top-k}, \\
   0 & \text{otherwise.}
   \end{cases}
\end{equation}

\textbf{Nucleus Sampling} \cite{Holtzman} involves choosing from a smaller set of plausible candidates by dynamically selecting a variable-sized subset of tokens (the "nucleus") that cumulatively make up a certain probability mass (e.g., top 90\%).
\begin{equation}
    P'(w) = 
   \begin{cases} 
   P(w \mid w_{<t}) & \text{if } \sum_{w' \in S} P(w' \mid w_{<t}) \leq p, \\
   0 & \text{otherwise.}
   \end{cases}
\end{equation}

We also implement two baselines, representing the execution-based sampling approaches \cite{codet, MBR, planning} and the post-revising approaches \cite{self_refine, Self-Edit,debug}, to demonstrate the efficiency of \ourapproach, Specifically,

\textbf {Sampling+Filtering} utilizes LLMs to generate a vast number of codes, which are then filtered by executing test cases.

\textbf{Post-revising} conducts testing after code is generated by LLMs, and further revises codes that fail these tests based on error messages.

\begin{table*}[ht]
\centering
\caption{The comparison of \ourapproach and baseline approaches on different code generation benchmarks. The \textbf{bold text} indicates the highest value for a particular metric within a given dataset, regardless of the baseline or its configurations.}
\label{main_result}
\resizebox{0.99\textwidth}{!}{
\begin{tabular}{lccccccccc}
\toprule
Approaches & \multicolumn{3}{c}{HumanEval (ET)} & \multicolumn{3}{c}{MBPP (ET)} & \multicolumn{3}{c}{CodeForces2305} \\ \cmidrule(r){2-4} \cmidrule(r){5-7} \cmidrule(r){8-10}
 & \multicolumn{1}{c}{PassRate} & \multicolumn{1}{c}{AvgPassRate} & \multicolumn{1}{c}{CCP} & \multicolumn{1}{c}{PassRate} & \multicolumn{1}{c}{AvgPassRate} & \multicolumn{1}{c}{CCP} & \multicolumn{1}{c}{PassRate} & \multicolumn{1}{c}{AvgPassRate} & \multicolumn{1}{c}{CCP} \\ \midrule
PG-TD & 46.3 (38.4) & 64.7 (61.2) & 82.3 & - & - & - & 0.0 & 0.0 & 6.7 \\
MGD & 31.4 (24.6) & 57.7 (54.3) & 79.5 &  34.9 (28.0) &  39.8 (40.4) & 82.4 & 0.0 & 0.0 & 0.0 \\
MBR-EXEC & 34.8 (28.1) & 59.3 (54.7) & 80.5 & 34.0 (27.6) & 39.0 (40.0) & 83.5 & 0.0 & 0.7 & 5.6 \\
AdapT & 31.9 (26.7) & 59.2 (55.8) & 78.2 & 32.4 (26.4) & 38.6 (39.1) & 80.9 & 0.0 & 0.1 & 5.6 \\
\hdashline Temperature Sampling ($T = 0.0$) & 31.1 (24.4) & 57.6 (54.5) & 80.5 & 35.5 (29.3) & 41.4 (42.0) & 82.0 & 0.0 & 0.0 & 6.7 \\
+ Post-revising & 31.3 (24.5) & 57.5 (54.5) & 79.9 & 35.7 (29.4) & 41.6 (42.0) & 82.1 & 0.0 & 0.0 & 6.7 \\
\rowcolor[gray]{0.9} \ourapproach & 53.8 (45.5) & 66.3 (62.6) & 95.8 & \textbf{40.5} (32.3) & \textbf{46.7} (47.3) & 98.7 &  8.3  &  21.2 & 85.9 \\
Temperature Sampling ($T = 0.6$) & 27.1 (22.9) & 52.2 (47.2) & 74.9 & 29.1 (23.3) & 34.7 (35.7) & 79.1 & 0.0 & 0.5 & 5.5 \\
+ Filtering & 35.9 (30.5) & 55.5 (51.5) & 74.4 & 31.8 (25.4) & 38.4 (38.9) & 88.1 & 0.0 & 1.8 & 6.7 \\
+ Post-revising & 32.3 (27.0) & 57.0 (51.8) & 75.7 & 30.2 (23.9) & 36.0 (36.8) & 83.9 & 0.0 & 0.4 & 4.3 \\
\rowcolor[gray]{0.9} \ourapproach & 55.8 (48.6) & 71.4 (66.7) & 96.4 & 35.7 (27.8) & 41.9 (42.4) & 98.3 &  \textbf{8.9} & 22.3 & \textbf{88.7} \\
Temperature Sampling ($T = 0.8$) & 22.2 (18.9) & 43.7 (38.4) & 67.9 & 21.9 (17.6) & 26.5 (27.3) & 69.3 & 0.0 & 0.6 & 4.4 \\
+ Filtering & 31.7 (29.9) & 51.7 (47.2) & 72.6 & 24.2 (20.1) & 30.0 (30.5) & 85.6 & 0.0 & 0.6 & 4.4 \\
+ Post-revising & 29.6 (25.9) & 51.8 (47.0) & 70.1 & 22.9 (18.1) & 29.4 (29.8) & 84.1 & 0.1 & 0.3 & 3.2 \\
\rowcolor[gray]{0.9} \ourapproach & 54.4 (48.0) & 70.9 (66.8) & \textbf{98.2}  & 30.3 (23.9) & 39.5 (40.9)  & 97.9 & 8.4 & 21.4 & 86.8  \\
\hdashline Top-k Sampling ($k=10$) & 22.3 (18.5) & 45.0 (40.1) & 69.3 & 23.2 (18.4) & 27.7 (28.5) & 70.9 & 0.0 & 3.6 & 5.9 \\
+ Filtering & 35.4 (30.5) & 55.9 (51.2) & 68.9 & 25.4 (20.4) & 30.3 (31.6) & 85.4 & 0.0 & 0.4 & 7.8 \\
+ Post-revising & 27.1 (22.7) & 50.1 (45.6) & 70.5 & 25.5 (20.1) & 30.5 (31.3) & 83.1 & 0.0 & 0.6 & 5.4 \\
\rowcolor[gray]{0.9} \ourapproach & 53.1 (39.7) & 67.7 (65.0) & 97.0 & 32.6 (25.2) & 40.7 (41.6)  & 97.6 & 7.7 & 20.1 & 86.6  \\
Top-k Sampling ($k=40$) & 21.7 (19.0) & 43.8 (39.2) & 70.4 & 22.3 (17.7) & 26.8 (27.6) & 69.3 & 0.0 & 0.3 & 5.0 \\
+ Filtering & 33.5 (29.3) & 56.4 (51.6) & 78.0 & 25.4 (20.2) & 30.8 (31.2) & 84.8 & 0.0 & 0.0 & 4.4 \\
+ Post-revising & 29.3 (25.2) & 52.3 (47.4) & 69.1 & 24.5 (19.3) & 29.5 (30.4) & 82.1 & 0.1 & 0.4 & 4.4 \\
\rowcolor[gray]{0.9} \ourapproach & 55.0 (46.1) & 67.6 (62.7) & 96.4 & 31.5 (24.7) & 39.8 (41.4)  & 98.1 & 8.1 & 20.8 & 87.2  \\
\hdashline Nucleus Sampling ($p=0.8$) & 28.8 (23.4) & 54.7 (49.4) & 76.3 & 31.2 (25.7) & 37.0 (37.7) & 81.2 & 0.0 & 5.2 & 6.1 \\
+ Filtering & 34.8 (28.1) & 63.4 (57.5) & 78.7 & 31.9 (24.7) & 38.5 (39.3) & 88.8 & 0.0 & 0.8 & 4.4 \\
+ Post-revising & 30.0 (24.2) & 56.1 (50.8) & 75.3 & 31.9 (25.2) & 37.7 (38.3) & 84.1 & 0.1 & 0.5 & 5.6 \\
\rowcolor[gray]{0.9} \ourapproach & 55.3 (46.8) & 71.2 (68.0) & 98.0  & 38.4 (30.5) & 44.7 (45.6) & 98.5 & 8.3  &  21.9 & 87.9 \\
Nucleus Sampling ($p=0.9$) & 26.7 (22.2) & 52.0 (47.5) & 75.1 & 28.5 (22.5) & 33.9 (35.0) & 79.3 & 0.0 & 0.4 & 6.0 \\
+ Filtering & 36.0 (30.5) & 59.2 (54.9) & 79.8 & 31.0 (24.1) & 36.7 (37.5) & 89.4 & 0.0 & 0.4 & 5.6  \\
+ Post-revising & 32.5 (27.2) & 58.0 (53.8) & 76.3 & 29.4 (23.1) & 34.8 (35.7) & 82.8 & 0.1 & 0.4 & 4.9 \\
\rowcolor[gray]{0.9} \ourapproach & \textbf{57.3} (48.2) & \textbf{72.4} (66.1) & 97.6 & 36.1 (27.5) & 42.1 (42.6) & \textbf{99.1}  & 8.7 & \textbf{23.4} & 88.2 \\ \bottomrule
\end{tabular}}

\end{table*}

Additionally, we also compare four state-of-the-art (SOTA) code generation approaches that operate during the decoding process, Specifically,

\textbf{PG-TD} \cite{planning} employs Monte Carlo Tree Search during the LLM decoding process, formulating rewards based on testing results to guide the generation of code.

\textbf{MBR-EXEC} \cite{MBR} introduces the execution result-based minimum Bayes risk decoding to select code from the samples generated by LLMs.

\textbf{MGD} \cite{monitor} utilizes static analysis tools to perform type analysis at pre-defined trigger points (specifically at dereference operations) during the code generation process of LLMs, enabling the selection of type-consistent variables.

\textbf{AdapT} \cite{adapt} dynamically adjusts the temperature during the LLMs' generation process, applying a higher temperature at points of low generation probability (challenging tokens) and a lower temperature at points of high generation probability (confident tokens).
\\
\\

\noindent \textit{\textbf{3) Metrics:}}
We used three metrics to evaluate our approach, including PassRate, AvgPassRate, and Compiler Correctness Percentage.

\textbf{PassRate} \cite{codex} metric can measure the functional correctness of the generated code by executing private test cases. For each task, $n \geq 1$ samples of code are generated, and the number of samples, $c \leq n$, that pass the test cases are counted. The PassRate is then calculated using the following estimator:
\begin{equation}
    \operatorname{PassRate} = \mathop{\mathbb{E}}\limits_{\operatorname{Problems}}\begin{bmatrix}1-\tiny{\frac{\begin{pmatrix}n-c\\1\end{pmatrix}}{\begin{pmatrix}n\\1\end{pmatrix}}}\end{bmatrix}.
\end{equation}

\textbf{AvgPassRatio} \cite{apps} calculates the average proportion of test cases that generated codes $\mathbf{y}_p's$ pass, which is a milder metric than PassRate, allowing to assess the partial correctness of the generated codes. 
\begin{equation}
    \label{AvgPassRatio}
    \frac{1}{|P|} \sum_{p\in P} \frac{1}{|C_{p}|} \sum_{c\in C_{p}} \mathbb{I}\left\{\operatorname{Eval}\left(\mathbf{y}_p, \mathcal{I}_{p,c} \right)=\mathcal{O}_{p,c}\right\},
\end{equation}
where $ p $ represents a task within the test set $ P $, and $ \{(\mathcal{I}_{p,c}, \mathcal{O}_{p,c})\}_{c=1}^{C_p} $ is the set of test cases for $ p $, $\mathbb{I}(\cdot)$ is an indicator function, which outputs 1 if the condition is true and 0 otherwise, and $\operatorname{Eval}\left(\mathbf{y}_p, \mathcal{I}_{p,c} \right)$ represents an evaluation function that obtains outputs of code $\mathbf{y}_p$ by way of executing it with $\mathcal{I}_{p,c}$ as input. 

\textbf{Compiler Correctness Percentage (CCP)} measures the proportion of generated code samples that are compilable (i.e., free of syntax errors and compilation errors). It is defined as:

\begin{equation}
\text{CCP} = \frac{N_{\text{compilable}}}{N_{\text{total}}},
\end{equation}
where $ N_{\text{compilable}} $ is the number of compilable code samples, and $ N_{\text{total}} $ is the total number of generated codes.
\\
\\
\noindent \textit{\textbf{4) Implementation Details:}} In the evaluation, we use CodeLlama-7B \cite{CodeLlama}  as base model by default. The decay factor $\lambda$ for constraint regeneration is set at 0.9. The maximum generation length of our approach and baselines is set to 512 on all benchmarks, except for CodeForces2305, where it is set to 1024. To mitigate the instability of the model sampling, we report the average results of three trials in the experiments. Due to space limits, we only present the results on the HumanEval dataset (other benchmarks follow similar trends) for RQ3, RQ4, and RQ5.

\subsection{RQ1. Comparing \ourapproach to Baseline Approaches}
To evaluate the effectiveness of \ourapproach on code generation, we evaluate test correctness and compile correctness across various representative code generation benchmarks, including HumanEval, MBPP, HumanEval-ET, MBPP-ET, and CodeForces2305.

\indent \textbf{Settings:}
We compare our approach with nine baselines, including Temperature Sampling, Topk Sampling, Nucleus Sampling, Sampling+Filtering, Post-revising, PG-TD, MBR-EXEC, MGD, and AdapT. For the HumanEval, MBPP, HumanEval-ET, and MBPP-ET benchmarks, CodeLlama-7B serves as our base model, while for the more challenging CodeForces2305 benchmark, we employ CodeLlama-34B as our base model. Since Temperature Sampling, Top-k Sampling, and Nucleus Sampling are sensitive to their parameter settings, we evaluate their performance under different settings. For the temperature (T) in Temperature Sampling, we use values of 0.0, 0.6, and 0.8. For the k value in Top-k Sampling, we use 10 and 40. For the p-value in Nucleus Sampling, we use 0.8 and 0.9. Our approach, Sampling+Filtering, and Post-revising can be combined with these three decoding methods. We set the token budget of \ourapproach during generation to be twice the maximum generation length. Sampling+Filtering and Post-revising maintain the same token budget as \ourapproach.

\indent \textbf{Results:}
The experimental results are shown in Table \ref{main_result}. 
These results demonstrate that our approach outperforms all baseline approaches across three metrics on five datasets, demonstrating the superior performance of \ourapproach. Notably, our approach shows the best performance at 57.3\% in pass rate under the Nucleus Sampling ($p = 0.9$) setting on HumanEval benchmark, exceeding the direct generation with LLMs by 30.6\% in the same setting. Specifically, our approach exceeds those of Sampling+Filtering and Post-revising across three commonly used decoding methods: Temperature Sampling, Top-k Sampling, and Nucleus Sampling. The fact that our approach significantly surpasses the Sampling+Filtering approach proves that the improvement in performance is not merely due to repetitive sampling but is greatly aided by the backtracking mechanism. Compared to Post-revising, our approach can resolve errors in real-time during the generation process, which helps enhance the quality of generated code and prevents the accumulation of errors that can complicate error resolution. Among all baselines, PG-TD performs the best; however, it requires both test inputs and outputs for execution, limiting its applicability to benchmarks like MBPP that do not provide public test cases. It is also worth noting that all approaches generally perform worse on the CodeForces2305 dataset compared to other benchmarks. This may be due to two reasons: firstly, the code generation task in CodeForces2305 is inherently challenging, with even the powerful ChatGPT achieving only a 7.9\% pass rate in the original paper \cite{dong-etal-2024-generalization}; secondly, potential data contamination issues might have caused the LLMs to perform exceptionally well on other benchmarks, creating a significant disparity with CodeForces2305. Despite this, \ourapproach successfully attains a performance level of pass rate up to 8.9\%, which represents a substantial improvement over the baselines on the CodeForces2305 benchmark. This enhancement underscores the significant potential of our approach to elevate the capabilities of LLMs in addressing complex problem-solving tasks.

\subsection{RQ2. Performance on Multilingual Code Generation}
For different programming languages, due to the unique characteristics of each language and the distribution of training data, there are variations in the performance of LLMs when generating code in different languages. In this evaluation, we examine the performance of our approach on multilingual code generation tasks.

\indent \textbf{Settings:} In addition to Python language, we also evaluate our approach on C++ language utilizing HumanEval-CPP \cite{codegeex} benchmark. The baseline approaches include Temperature Sampling, Sampling + Filtering, and Post-revising, all of which employ the best-performing configurations of PassRate as shown in Table \ref{main_result}.

\begin{table}[h!]
\caption{The performance of \ourapproach on different programming languages (PL).}
\resizebox{0.46\textwidth}{!}{
\begin{tabular}{llccc}
\toprule
\multicolumn{1}{l}{PL} & Approaches & \multicolumn{1}{c}{PassRate} & \multicolumn{1}{c}{AvgPassRate} & \multicolumn{1}{c}{CCP} \\
\midrule
\multirow{3}{*}{C++} & Temperature Sampling & 26.8 & 45.2 & 85.8 \\
 & Sampling + Filtering &  29.5 & 49.8 & 84.6 \\
 & Post-revising & 28.9 & 48.8 & 85.4 \\
 & \cellcolor{gray!20}\ourapproach & \textbf{39.6}  & \textbf{60.7}  & \textbf{95.5} \\
\hdashline
\multirow{3}{*}{Python} & Temperature Sampling & 31.1 & 57.6 & 80.5 \\
 & Sampling + Filtering & 36.0 & 59.2 & 79.8 \\
 & Post-revising & 32.5 & 58.0 & 76.3 \\
 & \cellcolor{gray!20}\ourapproach & \textbf{57.3}  & \textbf{72.4}  & \textbf{97.6} \\
 \bottomrule
\end{tabular}} \label{PL_result}
\end{table}

\indent \textbf{Results:}
The experimental results in Table \ref{PL_result} show that our approach significantly improves performance in both languages. Our approach achieves greater improvement on Python, which is a language where LLMs excel, compared to C++. Nevertheless, our approach still outperforms all baselines in C++, with a relative increase of 34.2\% over the best-performing baseline, i.e., Sampleing + Filtering, in pass rate. Moreover, our approach achieves a 95.5\% compilation pass rate on C++ code generation tasks, significantly higher than other baselines. Utilizing compiler-based program analysis for error detection proves effective across various languages, ensuring the robustness and versatility of our approach.

\subsection{RQ3. Performance on Different LLMs}
\ourapproach is model-agnostic and can be applied to a variety of LLMs. In this evaluation, we explore how \ourapproach  enhances code generation performances across different LLMs.

\indent \textbf{Settings:}
We employ several different series and sizes of representative general LLMs and Code LLMs to perform \ourapproach. The general LLMs used are from the Llama series (Llama-2-7B, 13B, and 34B \cite{llama2}), while the Code LLMs include the multi-lingual CodeGen series (CodeGen-2B, 6B, and 16B \cite{codegen}), and the CodeLlama series (CodeLlama-7B, 13B, and 34B \cite{CodeLlama}).

\begin{figure}[h!]
    \centering
    \includegraphics[width=0.495\textwidth]{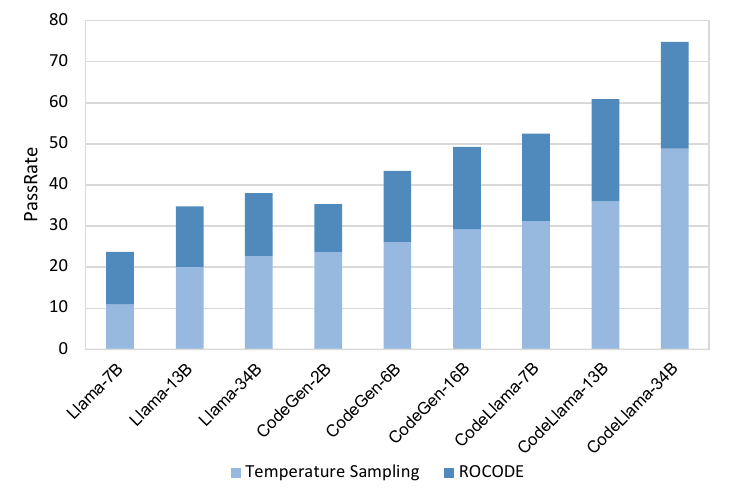}
    \caption{The performance of \ourapproach on different LLMs.}
    \label{diff_LLMs}
\end{figure}

\indent \textbf{Results:}
From the experimental results shown in Figure \ref{diff_LLMs}, we can observe that \ourapproach achieves significant improvements over temperature sampling across all series and on LLMs of various sizes. Our approach achieves higher performance on code LLMs compared to general LLMs, with a pass rate exceeding 70\%. Furthermore, we observed a trend in the enhancement across different LLMs: the stronger the base model, the greater the improvement brought by \ourapproach. This might suggest that more powerful LLMs have greater potential for enhancements through rollback corrections.

\subsection{RQ4. Cost and Efficiency of \ourapproach}

Besides performance,  cost, and efficiency also influence whether a code generation approach will be widely adopted. Therefore, we discuss the cost and efficiency of \ourapproach.

\indent \textbf{Settings:}
We measure the costs by the number of tokens consumed, since the computational resource usage for LLMs scales with the number of tokens and services that provide LLM access typically charge based on token usage. We also measure the efficiency by the speed of the running time (min). We compared \ourapproach with seven baseline methods: PG-TD, MGD, MBR-EXEC, AdapT, Temperature Sampling, Sampling+Filtering, and Post-revising. For Temperature Sampling, Sampling+Filtering, and Post-revising, we configure them according to the parameter configurations that exhibit the best performance (PassRate) shown in Table \ref{main_result}.

\begin{table}[h!]
\centering
\caption{The cost and efficiency of \ourapproach, where the \textbf{\textit{BOLD ITALIC}} indicates the highest value other than \ourapproach, which is also the baseline of the relative improvement.}
\resizebox{0.49\textwidth}{!}{
\begin{tabular}{lccc}
\toprule
Approaches & PassRate & Token Consumption & Time  \\
\midrule
PG-TD & \textbf{\textit{46.3}} & 675.2 & 1.219 \\
MGD & 31.4 & 566.4 & 0.348 \\
MBR-EXEC & 34.8 & 438.5 & 0.314 \\
AdapT & 31.9 & 332.1 & 0.309 \\
Temperature Sampling & 31.1 & 445.8 & 0.313 \\
Sampling + Filtering & 36.0 & 532.2 & 0.334 \\
Post-revising & 32.5 & 623.4 & 0.498 \\ \hdashline
\ourapproach & \textbf{57.3} \ \ (\textcolor{red}{ $\uparrow$ 23.8\%}) & 503.1 & 0.622 \\
\bottomrule
\end{tabular}} \label{cost_result}
\end{table}

\indent \textbf{Results:}
The evaluation results on HumanEval benchmark are presented in Table \ref{cost_result}. In terms of cost (token consumption), our approach shows clear advantages compared to most baselines. Compared to Temperature Sampling that is directly generated with LLMs, our cost increases by less than 1.1 times.
Notably, our approach is substantially more efficient compared to Post-revising, with a cost reduction of 19.3\%. More importantly, the cost of our approach is significantly lower than the state-of-the-art (SOTA) approach, PG-TD. In terms of time efficiency, although our approach is slightly slower than Sampling + Filtering and Post-revising, it is still faster than PG-TD. The additional time cost primarily stems from calling the compiler for incremental checks. Considering the generated code is in Python language, which is an interpreted language suitable for just-in-time execution and dynamic typing, we can perform incremental execution rather than starting from scratch each time to further optimize the speed of checks.

\subsection{RQ5. Ablation Study}

\ourapproach consists of three key components: incremental error detection, strategic rollback, and constraint generation. We evaluate the effectiveness of each component through ablation experiments.

\indent \textbf{Settings:}
We modify or remove different components while keeping the rest of \ourapproach unchanged:
\begin{itemize}
    \item For error detection, we replace the original program analysis-based detection with an entropy-based detection, which aligns with our rollback strategy, reporting errors at locations with the highest entropy (\textbf{Entropy-based Error Detection}).
    \item For the strategic rollback, we explore four other rollback strategies respectively:
      1) Roll back directly to the beginning and generate from scratch (\textbf{Full Restart Rollback}).
      2) Roll back only to the statement of the reported error (\textbf{Error Statement Rollback}).
      3) Roll back only to the statement with the highest entropy (\textbf{High Entropy Statement Rollback)}).
      4) Roll back to the token with the highest entropy and disable that token (\textbf{High Entropy Token Disable Rollback}).
  \item For constraint generation, we remove the constraints during generation but instead resample an output (\textbf{Constraint-Free Resampling}).
\end{itemize}

\begin{table}[h!]
\centering
\caption{Ablation results.}
\resizebox{0.49\textwidth}{!}{
\begin{tabular}{lccc}
\toprule
Variants & PassRate & AvgPassRate & CPP \\
\midrule
Entropy-based Error Detection & 45.1 & 59.0 & 76.2 \\
Full Restart Rollback & 50.6 & 63.8 & 95.0 \\
Error Statement Rollback & 51.2 & 62.9 & 93.2 \\
High Entropy Statement Rollback & 49.4 & 61.9 & 92.8 \\
High Entropy Token Disable Rollback & 47.5 & 60.7 & 88.3 \\
Constraint-Free Resampling & 50.7 & 64.4 & 89.4 \\
\hdashline
\ourapproach & \textbf{57.3} & \textbf{72.4} & \textbf{97.6} \\

\bottomrule
\end{tabular}} \label{abaltion}
\end{table}

\indent \textbf{Results:}
The experimental results on the HumanEval benchmark are shown in Table \ref{abaltion}. From the experimental results, it is evident that all components of our approach are effective. In contrast to the entropy-based error detection approach, our program analysis-based error detection avoids the bias of labeling tokens as erroneous merely due to their high entropy, as not all high-entropy tokens lead to errors. The Full Restart Rollback cannot achieve the same performance as our approach under the same token budget as it lacks in efficiency. The performance decline observed in both Error Statement Rollback and High Entropy Statement Rollback further validates the effectiveness of combining program analysis with LLM-based entropy assessments in rollback strategies. Additionally, approaches that simply block high-entropy tokens have failed to effectively alter the entropy of the code, thus offering limited performance enhancement. Removing constraints during decoding also leads to a noticeable decline in performance, which confirms the efficacy of our Constraint Regeneration approach.

\subsection{RQ6. Effect of Decay Factor}

Since \ourapproach involved one hyperparameter, the decay factor $\lambda$, we evaluate the impact of different values of this hyperparameter on the performance to analyze its sensitivity.

\indent \textbf{Settings:}
We choose \{0.5, 0.6, 0.7, 0.8, 0.85, 0.9, 0.95, 0.98, 0.99\} as test values for $\lambda$. We conduct experiments on HumanEval benchmarks under the setting of greedy sampling.

\begin{figure}[th!]
    \centering
    \includegraphics[width=0.49\textwidth]{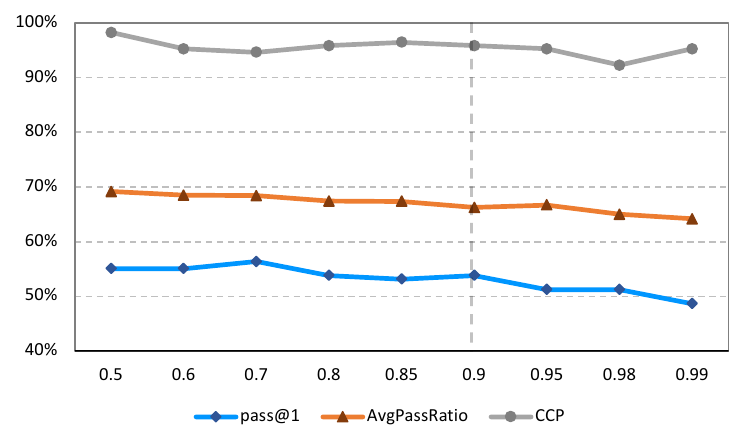}
    \caption{The performance of \ourapproach with different values of the hyperparameter $\lambda$. We use the \textcolor{gray}{gray dashed line} to represent the employed hyper-parameters.}
    \label{decay_factor}
\end{figure}

\indent \textbf{Results:}
The experimental results of this evaluation are shown in Figure \ref{decay_factor}. From the results we can observe that as the hyperparameter $\lambda$ increases, the metrics PassRate and AvgPassRate show a slight downward trend, although the decline is not significant. On the other hand, the CCP metric, while fluctuating across different values of r, still maintains a high level overall, averaging over 90\%. These observations suggest that our approach demonstrates strong robustness to adjustments in the hyperparameter $\lambda$. The downward trend in PassRate and AvgPassRate could be due to higher values of $\lambda$ meaning looser constraint penalties during code generation, which decays the likelihood of previous errors less. Specifically, a larger $\lambda$ value reduces the immediate penalty for errors, requiring more iterations to correct mistakes, which may affect the performance of generating correct code within a limited token budget.

\section{Threats to Validity}

There are three major threats to the validity of our work. 

\textbf{1) Threats to external validity} concern the quality of experimental datasets and the generalizability of our results. First, we use six public code generation datasets for evaluation, which are mainstream benchmarks and have been used in many related works \cite{wizardcoder, starcoder, dataset_previous_1,dataset_previous_2,planning,dataset_previous_4}. Moreover, to prevent the evaluation dataset from being affected by data contamination (i.e., the test data may have been included in the training data of LLMs), we used problems from CodeForces that were published after the cutoff date of LLM‘s training data for the assessment. Second, \ourapproach can be applied to any LLMs, and we choose nine well-known LLMs \cite{model_previous_1,model_previous_2,model_previous_3,model_previous_4} of different series and sizes for our experiments. 
    
\textbf{2) Threats to internal validity} 
involve the impact of hyperparameters. For our approach, we introduce a hyperparameter, i.e. the decay factor in constrained regeneration. For this hyperparameter, we intuitively selected a specific value and observed that it enhances performance across multiple benchmarks. To further explore the impact of this hyperparameter, we conducted detailed experimental studies, which showed that this hyperparameter effectively improves experimental results over a broad range. As for other hyperparameters, such as maximum generation length and temperature, to ensure fairness in comparison, we maintained these parameters consistent with the baseline approaches. 

\textbf{3) Threats to construct validity} pertain to the reliability of evaluation metrics. We use the test pass rate as the primary evaluation metric. However, due to the limited number of test cases, this method cannot fully assess the functional correctness of the generated code. To mitigate this issue, we adopted extended versions of some benchmarks, which significantly expanded the number of test cases to provide a more comprehensive functional evaluation. For PassRate metrics, we employ the unbiased version of PassRate \cite{codex} to diminish evaluation errors that arise from sampling. On this basis, each experiment is run three times, and its average result is reported.

\section{Related Work}

\subsection{Code Generation with LLMs}

General LLMs, represented by ChatGPT \cite{ChatGPT}, have demonstrated significant potential in software engineering tasks such as code generation. This led to the development of specialized LLMs for code generation, such as AlphaCode \cite{alphacode}, CodeGen \cite{codegen}, Incoder \cite{incoder}, CodeGeeX \cite{codegeex}, Starcoder \cite{starcoder}, WizardCoder \cite{wizardcoder} and CodeLlama \cite{CodeLlama}. These specialized LLMs are typically developed by further training general LLMs or by training them from scratch using code corpus. Furthermore, there is a series of research efforts for code generation that propose improvements to the decoding stage of general LLMs or Code LLMs. Zhang et al. \cite{planning} proposed a planning-guided decoding algorithm to generate higher-quality programs. This algorithm is based on Monte Carlo Tree Search (MCTS) and explores different branches of the search tree to examine various possibilities for program generation. After generating a complete program, it is evaluated by executing test cases to obtain rewards. Shi et al. \cite{MBR} and Chen et al. \cite{codet} generate a large number of program samples from LLMs and subsequently re-ranking them using public or generated test cases.
Zhang et al. \cite{Self-Edit} introduced Self-edit, which involves training another model to modify the programs generated by LLMs based on the results of executing test cases. Similarly, Cheng et al. \cite{debug} also developed a post-processing technique for modifying the outputs of models. Those approach leverages the capabilities of LLMs to debug and correct their own errors.

\subsection{Combining Program Analysis and LLMs}
Combining emerging LLMs with traditional program analysis techniques to overcome existing technological limitations has become a new trend. Currently, there have been some efforts in this direction, which have been applied to various software engineering tasks including program synthesis, formal verification, and defect detection. Jain et al. \cite{jigsaw} proposed Jigsaw, an approach that performs several transformations and checks during the processing steps, thereby enhancing the program synthesis capabilities of LLMs and validating it through the synthesis of the Python Pandas API. Agrawal et al. \cite{monitor} incorporated type-based static analysis into the code generation process, enabling the provision of a candidate list that constrains LLMs to produce type-correct identifiers. Wen et al. \cite{specification_synthesis} utilize static analysis techniques to decompose programs, thereby facilitating incremental specification generation for program verification. Li et al. \cite{UBI_bug} designed LLift, a framework that enables interaction between static analysis tools and LLMs, using use-before-initialization (UBI) bugs as a case study to demonstrate its effectiveness. Wang et al. \cite{inferROI} proposed a resource leak detection approach that combined LLMs with static program analysis. This approach utilizes LLMs to infer the resource-oriented intentions (resource acquisition, release, and reachability verification) in code, instead of matching predefined APIs, and then inferred intentions are applied to enhance static resource leak detection techniques.

\section{Conclusion}
In this paper, we introduce \ourapproach, a novel code generation approach based on LLMs that integrates backtracking mechanism and program analysis tools to eliminate errors in the code generation process. Our approach enables LLMs to generate programs incrementally, followed by incremental error detection through program analysis. When an error is detected, we perform rollback strategies, which provide an opportunity for LLMs to make modifications during the generation process. Furthermore, we impose constraints on the regeneration process to avoid repeating historical errors. Our approach is model-agnostic and does not require training, allowing for direct integration with LLMs. Experimental results show that our approach consistently outperforms baselines across various benchmarks, providing stable improvements for different decoding approaches and various LLMs.

\bibliographystyle{IEEEtran}

\bibliography{ref}

\newpage
\onecolumn
\appendix

\subsection{Examples of \ourapproach vs. Post-revising}

\begin{figure*}[h!]
    \centering
    \includegraphics[width=1.01\textwidth]{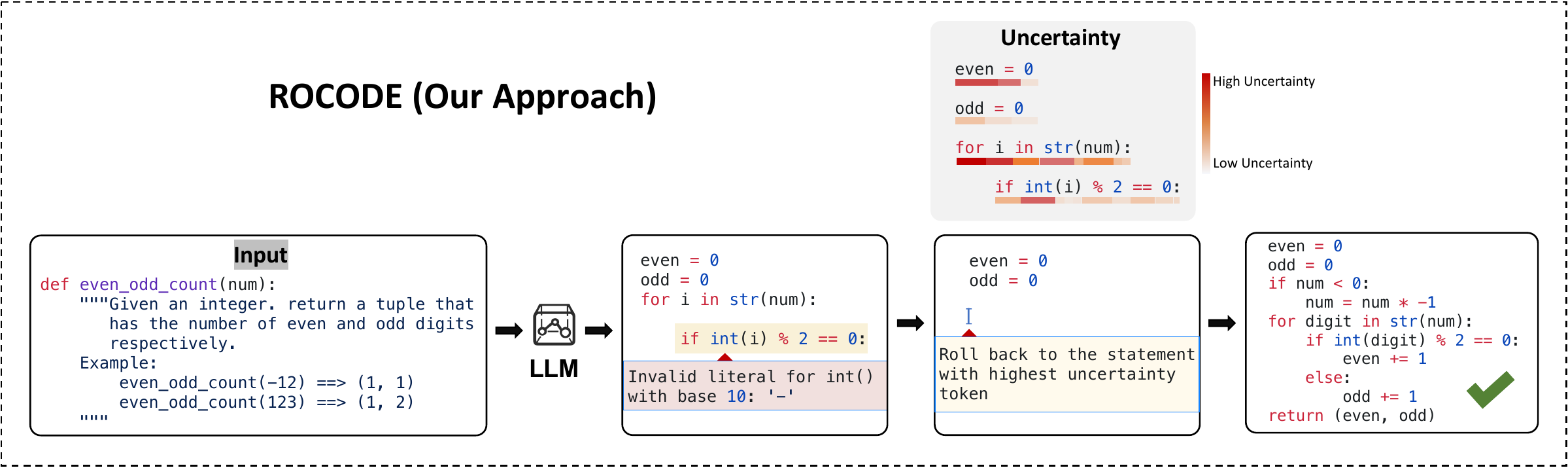}
    \caption{An example of \ourapproach.}
    \label{example_ROCODE}
\end{figure*}

\begin{figure*}[h!]
    \centering
    \includegraphics[width=0.8\textwidth]{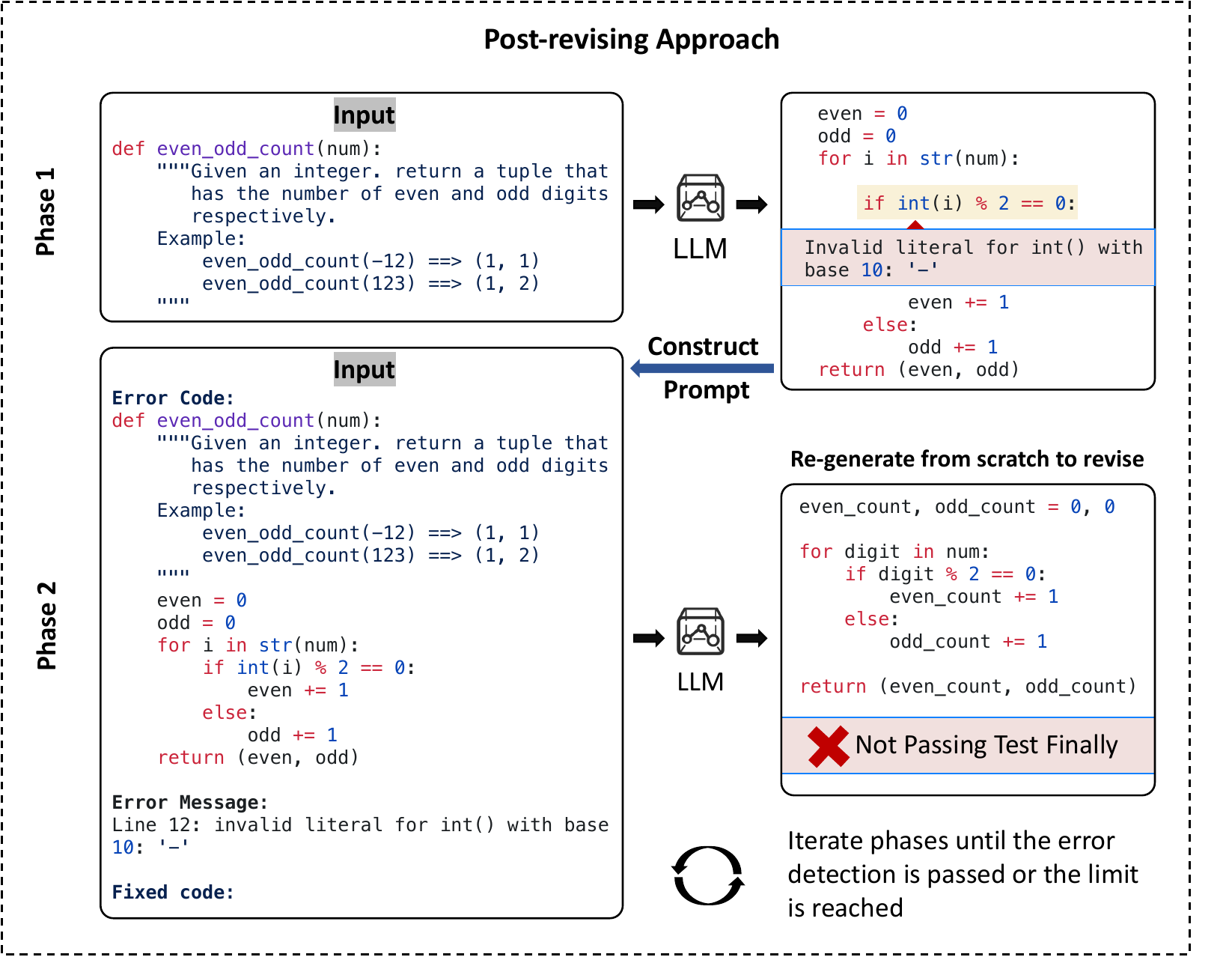}
    \caption{An example of Post-revising approach.}
    \label{example_Post-revising}
\end{figure*}

Figure \ref{example_ROCODE} presents an example of \ourapproach based on CodeLlama-7B, and Figure \ref{example_Post-revising} shows an example of the Post-revising approach with the same LLMs for comparison.

\end{document}